# Interplay of Quantum Size Effect and Tensile Strain on Surface Morphology of β-Sn(100) Islands

Bing Xia[1,2,6], Xiaoyin Li[3,6], Hongyuan Chen[1], Bo Yang[1], Jie Cai[1], Stephen Paolini[2], Zihao Wang[2], Zi-Jie Yan[2], Hao Yang[1,4,5], Xiaoxue Liu[1,4,5], Liang Liu[1,4,5], Dandan Guan[1,4,5], Shiyong Wang[1,4,5], Yaoyi Li[1,4,5], Canhua Liu[1,4,5], Hao Zheng[1,4,5], Cui-Zu Chang[2], Feng Liu[3], and Jinfeng Jia[1,4,5]

[1]TD Lee Institute and School of Physics and Astronomy, Shanghai Jiao Tong University, Shanghai 200240, China

[2]Department of Physics, The Pennsylvania State University, University Park, PA 16802, USA

[3]Department of Materials Science and Engineering, University of Utah, Salt Lake City, Utah 84112, USA

[4]Hefei National Laboratory, Hefei 230088, China

[5]Shanghai Research Center for Quantum Sciences, Shanghai 201315, China

[6]These authors contributed equally: Bing Xia and Xiaoyin Li

Corresponding authors: cxc955@psu.edu (C.-Z. C.); fliu@eng.utah.edu (F. L.); jfjia@sjtu.edu.cn (J. J.).

**Abstract:** The quantum size effect (QSE) and strain effect are two key factors influencing the surface morphology of thin films, which can increase film surface roughness through QSE-induced thickness oscillation and strain-induced island formation, respectively. Surface roughness usually manifests in the early stages of film growth and diminishes beyond a critical thickness. In this work, we employ molecular beam epitaxy (MBE) to grow β-Sn(100) islands with varying thickness $N$ on bilayer graphene-terminated 6H-SiC(0001) substrates. Scanning tunneling microscopy and spectroscopy measurements reveal an inverse surface roughness effect that highlights the interplay of QSE and misfit strain in shaping the surface **morphology of β-Sn(100) islands. For $N \leq 10$, the islands exhibit flat surfaces, while for $N \geq 26$, the island surfaces become corrugated and patterned. For the intermediate range, i.e., $12 \leq N \leq 24$, both flat and patterned surfaces coexist, with the percentage coverage of the patterned surface oscillating as a function of $N$. By performing density functional theory calculations, we demonstrate that the unusual surface pattern evolution in our MBE-grown β-Sn(100) islands is a result of the interplay between QSE-induced surface roughing and tensile strain-induced smoothening effect.**

**Main text:** As the thickness of a metal film approaches the electron Fermi wavelength, quantum confinement effect becomes pronounced. Confinement of electronic states gives rise to quantum size effects (QSE), leading to substantial modifications of the electronic band structure. Unlike the continuous bands in bulk materials, thin films exhibit discretized electronic states that form quantum well (QW) states (Fig. 1a), whose energy positions oscillate near the Fermi level as a function of film thickness $N$ (Ref. [1]). Within the free-electron model, the formation and successive filling of these quantized sub-bands cause many physical quantities of QW films to oscillate with $N$, before gradually converging to bulk values once confinement along the surface-normal direction is sufficiently weakened [2, 3]. These $N$-dependent oscillations, known as Friedel oscillations [2, 4, 5], can be further modified by long-period beating patterns and phase shifts [6-9]. One prominent consequence of band discretization is an oscillatory dependence of the film's total energy on $N$, in stark contrast to the linear $N$ dependence observed in thicker films [10].

The surface energy is generally defined as the difference between the total energy of an $N$-layer thin film and that of $N$ layers of bulk atoms, and thus quantifies the energy cost of creating a surface. Because the energy per atomic layer is approximately constant, the surface energy of ultrathin QW films oscillates with $N$. It gradually converges to a constant value as QSE diminishes in thicker films [11, 12]. Therefore, surface energy is a crucial metric for assessing surface stability. Together with the local work function and the energy position of the highest occupied QW states [9, 13-15], the oscillatory surface energy plays a key role in governing growth modes, thermal stability, and other QSE-related properties of metal thin films [2, 4, 6, 7, 9, 14, 16-19]. Specifically, $N$-dependent modulation of surface energy favors the formation of islands with energetically preferred thicknesses, leading to enhanced island density and increased surface roughness in the ultrathin regime.

Molecular beam epitaxy (MBE) is a technique for synthesizing thin films, heterostructures, and superlattices with near-atomic precision and exceptional purity. The metal thin films grown by MBE on semiconductor substrates can form a quasi-2D electronic system, offering opportunities to explore novel properties induced by QSE (Ref. [20]). On the other hand, when the thin metal films are grown on lattice-mismatched substrates, misfit strain inevitably arises. Together with QSE, strain effects can further enrich the growth behavior of these thin films, giving rise to novel properties. Over the past decades, the impact of misfit strain on thin film growth and surface morphology has been well studied in the classical regime [21, 22]. One well-known phenomenon is the compressive strain-induced classical surface roughness effect (SRE) (Fig. 1b), which is a typical manifestation of Asaro-Tiller-Grinfeld (ATG) instability [23-25]. While the ATG instability can, in principle, occur under tensile strain, it most commonly arises under compressive strain. From this perspective, the influence of QSE on surface morphology can be viewed as a $N$-dependent quantum SRE. In addition, quantum electronic stress, i.e., QSE-induced surface stress oscillation, which arises from charge carriers rather than lattice strain, has been observed in quantum confinement systems [26, 27]. However, due to the challenges in measuring strain/stress and the lack of compelling experimental evidence, our understanding of the interplay between QSE and misfit strain remains incomplete.

In this work, we employ MBE to grow β-Sn(100) islands with varying $N$ on bilayer graphene-terminated 6H-SiC(0001) substrates. The β-Sn(100) islands are expected to exhibit complex QSE-induced properties due to their intricate electronic band structure [28] and unique growth mode [29, 30]. A key feature of Sn growth on hexagonal substrates is a thickness-dependent phase transition from the α to β phases (Supplementary Fig. 1). Through *in situ* scanning tunneling microscopy and spectroscopy (STM/S) measurements, we observe an unusual and counterintuitive evolution of surface growth morphology. The islands show flat surfaces for $N \leq 10$ and patterned surfaces for $N \geq 26$. For the intermediate range, i.e., $12 \leq N \leq 24$, both flat and patterned surfaces coexist, with the percentage coverage of the patterned surface (PCPS) oscillating as a function of $N$. Our density functional theory (DFT) calculations reveal that the interplay between the QSE-induced quantum SRE and the tensile strain-induced inverse SRE (ISRE) drives this unusual surface pattern evolution in epitaxial β-Sn(100) islands.

Unlike the well-studied Pb(111) islands [4, 6-9, 14, 15, 27, 31-43], the QSE in β-Sn(100) islands is more complex because their band structure is much closer to the Fermi energy [44-46]. Specifically, bulk β-Sn has two intersections between electronic bands and the Fermi energy along the [100] direction, which can cause a distinct numerical relationship between the β-Sn(100) layer thickness (monolayer thickness $d_0 = a/2 = 2.97$ Å) and Fermi wavelength (i.e., $k_{F1} = 0.4714$ Å$^{-1}$ and $k_{F2} = 0.1925$ Å$^{-1}$). This relationship results in surface energy oscillations with periods of two layers [i.e., $\Delta N_1 = \pi/(k_{F1}d_0) = 2.24$] or five layers [$\Delta N_2 = \pi/(k_{F2}d_0) = 5.49$] (Supplementary Fig. 2). Therefore, the QSE in β-Sn(100) islands may induce more complex SRE that, in turn, changes their physical properties. Compared with Pb(111) film, which has a single band crossing the Fermi energy [14], the QSE-induced oscillations in β-Sn(100) film are relatively weak. As a result, they are insufficient to completely suppress the formation of the most unstable layers and manifest over a narrower thickness range, starting from a thicker initial layer. Moreover, the missing observations of $N = 11$, 25, 39, and 53 indicate a possible 14 ML beating pattern, but no evident beating pattern is achieved in our calculation results, and no phase reversal of even-odd oscillations is observed in our experiments after passing the potential nodal points. The above evidence indicates a weaker QSE in Sn films.

As noted above, misfit strain is inevitable during the MBE growth of thin metal films on lattice-mismatched substrates. The surface roughness of the film usually increases when the strain energy exceeds a critical threshold. Under compressive strain, the film surface develops ripple-like and/or island-like features beyond a critical $N$, which can lead to useful strain-induced self-assembly of quantum wires and quantum dots [47, 48]. This phenomenon has been known as compressive strain-induced classical SRE (Fig. 1b) (Ref. [22]). However, when the film has a patterned surface at equilibrium, applying sufficient tensile strain can flatten the film surface, which we refer to as ISRE (Fig. 1b). For the as-grown β-Sn(100) islands, tensile strain can be introduced during MBE growth and cooling processes because of the smaller lattice constant (Refs. [49-54]) and the larger thermal expansion coefficient of β-Sn(100) (Refs. [35, 55, 56]), compared to graphene (Supplementary Fig. 1). Moreover, the sharp metal-semiconductor interface between Sn and graphene suppresses inter-diffusion and chemical reactions [30]. Consequently, β-Sn(100) islands grown on a bilayer graphene-terminated 6H-SiC (0001) substrate offer a unique platform to study the interplay of QSE-induced quantum SRE and tensile-strain-induced classical ISRE because the QSE-induced quantum electronic stress is expected to couple with the misfit strain-induced classical surface stress. The weak QSE in Sn films is expected to exert a comparable influence on surface energy as the strain effect, enhancing the interplay between these two effects.

First, we perform STM/S measurements on β-Sn(100) islands. A large-scale STM image shows that isolated β-Sn(100) islands with low surface coverage are distributed on the terraces of the graphene substrate (Fig. 2a). Even for $N = 90$, where QSE is suppressed, the islands remain well isolated (Supplementary Fig. 3), consistent with spot-like diffraction features characteristic of 3D island growth in the RHEED pattern (Supplementary Fig. 4). The layer number $N$ is determined based on the monolayer thickness $d_0 \sim 2.97$ Å of β-Sn(100). For example, the measured height of the left β-Sn island along the red arrow in Fig. 2a is ~10.4 nm, corresponding to $N = 35$, whereas the right β-Sn island has a height of ~8.1 nm, corresponding to $N = 27$. By adjusting growth rates and duration, we obtain β-Sn(100) islands with $9 \leq N \leq 56$.

Atomic resolution STM images reveal two typical surface morphologies: a flat surface (Fig. 2b) and a patterned surface that emerges at specific $N$ (Figs. 2c to 2e). To more clearly quantify the evolution of the patterned surface area with $N$, we further acquire STM images with scan sizes comparable to the full lateral dimensions of individual β-Sn(100) islands (Figs. 2f to 2h). All STM images discussed here are acquired with $V_B < 1$ V. The patterned regions are defined as areas where every other atomic column is visible, with the atoms in these columns appearing more than 10 pm higher than those on the flat surface. We note that the precise height threshold depends on the STM tip condition (Supplementary Fig. 5). To characterize the surface morphology quantitatively, we define the PCPS as the ratio of the total patterned surface areas to the entire island area. For $N = 18$, the patterned area is ~43.1 nm$^2$ within a total island area of ~ 440 nm$^2$ (Fig. 2f), corresponding to a PCPS value of ~9.8%. For $N = 19$, the patterned area increases to ~174.0 nm$^2$ out of a total island area of ~366 nm$^2$ (Fig. 2g), yielding a PCPS value of ~47.5%. For $N = 20$, the patterned area decreases to ~33.9 nm$^2$ within the same total island area of ~366 nm$^2$ (Fig. 2h), giving a PCPS value of ~9.3%.

Through systematic data collection and statistical analysis (Supplementary Figs. 5 to 7), we derive a trend of PCPS as a function of $N$ (Fig. 3a). This trend is divided into four ranges: Range I, for $N = 9$ and $N = 10$, the films exhibit entirely flat surfaces without any patterned structures, i.e., PCPS ~ 0 (Figs. 2b and 3a). Range II, for $12 \leq N \leq 24$, patterned surfaces begin to emerge (Fig. 2c), and samples with odd $N$ display a greater PCPS than those with even $N$, confirmed by an apparent even-odd layer oscillation in corresponding PCPS values (Fig. 3a). Representative large-scale STM images of islands with $N = 18$, 19, and 20 (Figs. 2f to 2h) exhibit this behavior, where patterned surfaces are prominent for $N = 19$. In contrast, flat surfaces dominate for $N = 18$ and $N = 20$. Moreover, for $12 \leq N \leq 24$, the energy positions of the highest occupied QW state ($E_{HOQWS}$) and the lowest unoccupied QW state ($E_{LUQWS}$) also exhibit an apparent even-odd layer oscillation (Fig. 3b). Range III, for $26 \leq N \leq 32$, the PCPS value increases monotonically as $N$ increases (Fig. 2d). Range IV, for $N \geq 33$, patterned surfaces fully cover the entire film. Figure 2e shows the surface morphology of the β-Sn(100) island with $N = 80$, where no flat surfaces are observed. To further establish the presence of well-defined QW states, we systematically acquire STM spectra for β-Sn(100) islands with $9 \leq N \leq 56$. The energy spacing between $E_{LUQWS}$ and $E_{HOQWS}$ is found to scale inversely with $N$ (Supplementary Fig. 8), consistent with predications of QSE theory for Pb thin films [16] and providing strong evidence for robust quantum confinement in β-Sn(100) islands.

To investigate the physical mechanism underlying the unusual surface pattern evolution of β-Sn(100) islands, we employ DFT to calculate the surface energies as a function of $N$ (Refs. [14, 17, 34-36]) for both unstrained islands and islands under misfit strain. For each strain level considered, the misfit strain is fixed and applied uniformly across all layers, as imposed by the periodic boundary conditions. Guided by our high-resolution STM images of β-Sn islands on graphene-terminated 6H-SiC(0001) (Figs. 4b and 4d), our DFT calculations reveal two possible surface morphologies for β-Sn(100) islands: flat and patterned surfaces (Figs. 4a and 4c). The atomic structures of flat and patterned surfaces are slightly different. The flat surfaces exhibit a rectangular lattice, consistent with the (100) surface obtained by direct cleavage of the body-centered tetragonal structure of bulk β-Sn (Refs. [57]), in which all atoms in the same layer are located at the same height (i.e., the blue atoms in Fig. 4a). In contrast, the patterned surfaces display a height difference between adjacent atomic columns along [010] direction (i.e., the red atoms in Fig. 4c). For most cases examined in our DFT calculations, these relative height characteristics remain unchanged after structural relaxation, unless otherwise noted. Besides the distinct surface morphologies, our STM/S measurements reveal different lattice constants for flat and patterned surfaces. For the flat surfaces, the measured lattice constants are $b = 6.00 \pm 0.02$ Å and $c = 3.21 \pm 0.02$ Å, whereas for the patterned surfaces, $b = 5.90 \pm 0.02$ Å and $c = 3.18 \pm 0.02$ Å (Figs. 4b and 4d), independent of $N$ (Supplementary Fig. 9). This difference indicates the presence of misfit strain.

The equilibrium lattice constants of bulk β-Sn obtained from PBE calculations are $a = b = 5.94$ Å and $c = 3.21$ Å. We first calculate the surface energy evolution of β-Sn(100) islands at equilibrium by fixing the in-plane lattice constants to these bulk values. Free-standing β-Sn(100) films with $8 \leq N \leq 40$ are considered, and all atomic positions are fully relaxed. The resulting surface energies for flat and patterned surfaces are shown as hollow blue squares and red circles, respectively, in Fig. 5a. As $N$ increases, the surface energy of the patterned configuration converges to a lower value, indicating that patterned surfaces are energetically favored at equilibrium.

Given the known tendency of PBE to slightly overestimate lattice constants [58], and the close agreement between the calculated equilibrium lattice constants and those measured for the patterned surfaces, we infer that the patterned surfaces observed experimentally correspond to the equilibrium configuration. In contrast, the flat surfaces are likely under tensile strain. Therefore, we attribute the emergence of flat surfaces to an ISRE, in which tensile strain flattens the equilibrium patterned surface (Fig. 1b). Because the ratios $c/b$ are similar for both surface types, we consider biaxial tensile strain for simplicity. Based on the lattice constants determined from our STM/S measurements, the tensile strain is estimated as (6.00-5.90)/5.90 ≈ 1.7%. Therefore, we consider β-Sn(100) islands under biaxial tensile strains in the range of 1.6% ~ 2.0%. In our experiments, tensile strain gradually relaxes as $N$ increases. However, this continuous strain evolution cannot be captured in DFT slab calculations because periodic boundary conditions impose a fixed strain across all slab thicknesses. Instead, we evaluate surface energies at fixed strain levels for different $N$, specifically for biaxial tensile strains of ~1.6%, ~1.8%, and ~2.0%. We note that within this framework, the DFT results and experimental observations can only be compared indirectly by identifying consistent trends rather than through a one-to-one correspondence. Nevertheless, this approach provides meaningful insight into the interpretation of the experimentally observed surface evolution.

Our calculations show that for β-Sn(100) islands under ~2.0% tensile strain, most initially patterned surfaces relax into flat configurations (Fig. 5b), indicating that patterned surfaces are generally unstable at this strain level. This behavior is consistent with our experimental observations for thin β-Sn(100) islands with flat surfaces, where tensile strain is significant due to lattice mismatch with the substrate. Compared with the equilibrium case (Fig. 5a), the surface energy of flat surfaces under ~2.0% tensile strain (Supplementary Fig. 10) exhibits two notable differences: (*i*) the surface energy converges to a lower value, indicating enhanced stability of flat surfaces, and (*ii*) the QSE is strengthened, leading to larger oscillation amplitudes. The enhanced oscillation is consistent with the observations in Pb(111) films, where strain modifies the surface energy similarly [27]. We note that the odd-even layer oscillation observed in our MBE-grown β-Sn(100) islands closely matches our prediction of $\Delta N_1 = \pi/(k_{F1}d_0) = 2.24$ based on electronic band structure calculations (Supplementary Fig. 2). For β-Sn(100) islands under ~1.8% tensile strain, the DFT results show even closer agreement with our experiments (Fig. 5a, black curve). In this case, the surface energy of flat surfaces converges to a value lower than that of patterned surfaces at equilibrium (black and red curves in Fig. 5a, respectively). Moreover, for several intermediate thicknesses ($N$ = 15, 19, and 26; highlighted by green circles), initially flat surfaces relax into patterned morphologies. This behavior reflects the influence of QSE on the relative stability of flat and patterned surfaces at intermediate thicknesses under this strain level, consistent with our experimental observations in Range II. In contrast, for β-Sn(100) islands under ~1.6% tensile strain, thick islands with initially flat surfaces relax into patterned configurations, indicating that this strain level is insufficient to stabilize flat surfaces (Supplementary Fig. 10).

By combining our DFT calculations with our experimental observations, we propose that the unusual surface pattern evolution in β-Sn(100) islands (Fig. 3a) arises from the interplay between QSE and tensile strain. The significant lattice mismatch between the graphene substrate and Sn islands introduces substantial tensile strain when the β-Sn(100) islands are thin. For Range I, i.e., $N$ = 9 and $N$ =10 (Fig. 3a), our DFT calculations suggest that the flat surfaces of β-Sn(100) islands under tensile strain exhibit lower surface energies compared to the patterned surfaces of equilibrium films (Fig. 5a). As a result, flat surfaces are stable and dominant in Range I. As the β-Sn(100) islands become thicker, substrate-induced tensile strain diminishes, leading to the behaviors observed in Ranges II, III, and IV (Fig. 3a). For Range II, where the PCPS trend exhibits the most complex behavior, $N$ remains insufficient to relieve the graphene substrate-induced tensile strain fully. Therefore, our ~1.8% biaxial tensile strain calculations apply to Range II. The surface energy for flat surfaces shows a strong odd-even layer oscillation. When $N$ is an even number, the surface energy for flat surfaces is low, causing flat surfaces to dominate, while patterned surfaces are less likely to form. However, when $N$ is odd, the surface energy for flat surfaces becomes high, making the formation of flat surfaces energetically unfavorable compared to the patterned surfaces at the equilibrium lattice constant (Fig. 5a). Therefore, the surface forms unstrained patterned structures to achieve greater stability. The behavior observed in Range II can be directly linked to QW states: β-Sn(100) islands with even $N$ exhibit a lower LUQWS energy (Fig. 3b), resulting in a lower flat-surface energy than islands with odd $N$. Consequently, β-Sn(100) islands with even $N$ favor a larger fraction of flat surface and exhibit smaller PCPS values (Fig. 3a).

Moreover, a phase shift induced by the graphene substrate may occur [33], causing an odd-even oscillation inversion, but this does not affect the main conclusion of this work. More discussion of the effects of the substrate and the underlying α-Sn layers on the surface energy of β-Sn(100) islands can be found in Supporting Information. In Range III, the tensile strain is further relieved as $N$ increases, making the patterned surfaces increasingly stable. Consequently, the surface energy can drop below the low-energy value of the oscillating surface energy for flat surfaces, leading to surface morphology with progressively greater patterned coverage. Range IV represents a saturation of Range III, where the tensile strain has been fully relieved for large $N$. The patterned surfaces are energetically favorable for β-Sn(100) islands without tensile strain, consistent with the experimental observation of a fully patterned surface across the β-Sn(100) island. More discussion of surface pattern formation and the associated strain-relaxation mechanisms during the growth of β-Sn(100) islands can be found in Supporting Information.

To summarize, we employ MBE to grow β-Sn(100) islands with different thicknesses on bilayer graphene-terminated 6H-SiC(0001) substrates and observe an unusual evolution of surface morphology in these islands. Within a specific thickness range, we demonstrate a pronounced thickness-dependent evolution of the percentage coverage of patterned surfaces. In contrast to all previously studied systems [7, 16, 19, 44, 59], β-Sn(100) islands exhibit an increase in surface roughness with increasing thickness. Based on DFT calculations, we explain this unusual surface morphology evolution through the interplay between QSE and tensile strain. Our results demonstrate a gradual strain relaxation process as the QW films grow thicker and provide direct experimental evidence for the coexistence and interplay between quantum and classical effects manifested by the coupling between quantum electronic stress and classical misfit strain.

**Supporting Information.** The Supporting Information is available free of charge on the ACS Publications website.

Heat treatment of 6H-SiC(0001), MBE growth of β-Sn(100) islands, phase transition during Sn growth on graphene, extraction of PCPS values, more discussion of the effects of the graphene substrate and the underlying α-Sn layers on the surface energy of β-Sn(100) islands, additional theoretical calculations and STM/S results, more discussion on flat surface energies of β-Sn(100) films under different biaxial tensile strains, and the mechanisms of surface pattern formation, strain relaxation, and surface morphology.

**Author contributions:** J. J. conceived and supervised the experiment. B. X., H. C., B. Y., S. P., W. Z., Y. Z., and J. C. grew the β-Sn(100) islands and performed the STM/S measurements. X. L. and F. L. provided theoretical support. B. X., X. L., C.-Z. C., F. L. and J. J. analyzed the data and wrote the manuscript with input from all authors.

**Notes:** The authors declare no competing financial interest.

**Acknowledgments:** This work is supported by NSFC (11790313, 92065201, 11874256, 11874258, 12074247, 12174252, and 11861161003), the Ministry of Science and Technology of China (2019YFA0308600 and 2020YFA0309000), the Strategic Priority Research Program of Chinese Academy of Sciences (XDB28000000), the Science and Technology Commission of Shanghai Municipality (2019SHZDZX01, 19JC1412701, and 20QA1405100), and the Innovation Program for Quantum Science and Technology (2021ZD0302500). The theoretical work done at the University of Utah (X.Y. Li and F. Liu) is supported by the U.S. Department of Energy (DOE)-Basic Energy Sciences under Grant No. DE-FG02-04ER46148 and computational resources provided by the CHPC at the University of Utah and DOE-NERSC. The work done at Penn State is supported by the NSF grant (DMR-2241327) and the Gordon and Betty Moore Foundation's EPiQS Initiative (GBMF9063 to C. -Z. C.).

**Figures and figure captions:**

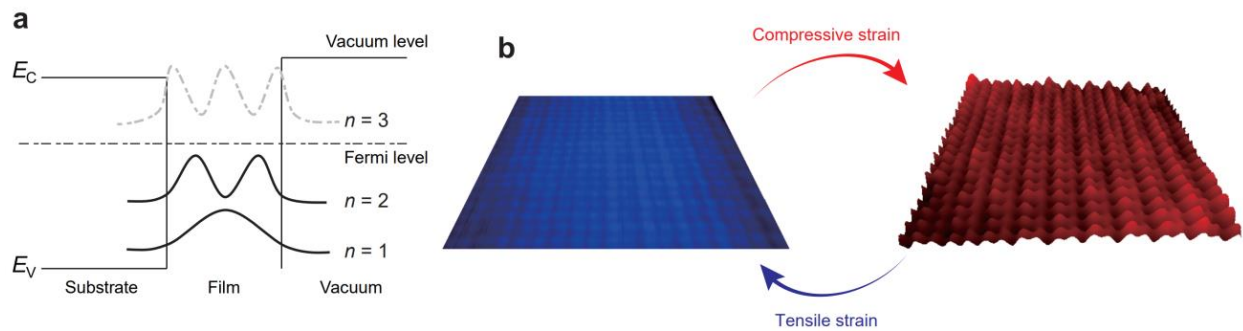

**Fig. 1| QSE and misfit strain in thin films. a,** Formation of QW states in thin films. **b,** Schematics of classical SRE induced by compressive strain (from left to right) and ISRE induced by tensile strain (from right to left).

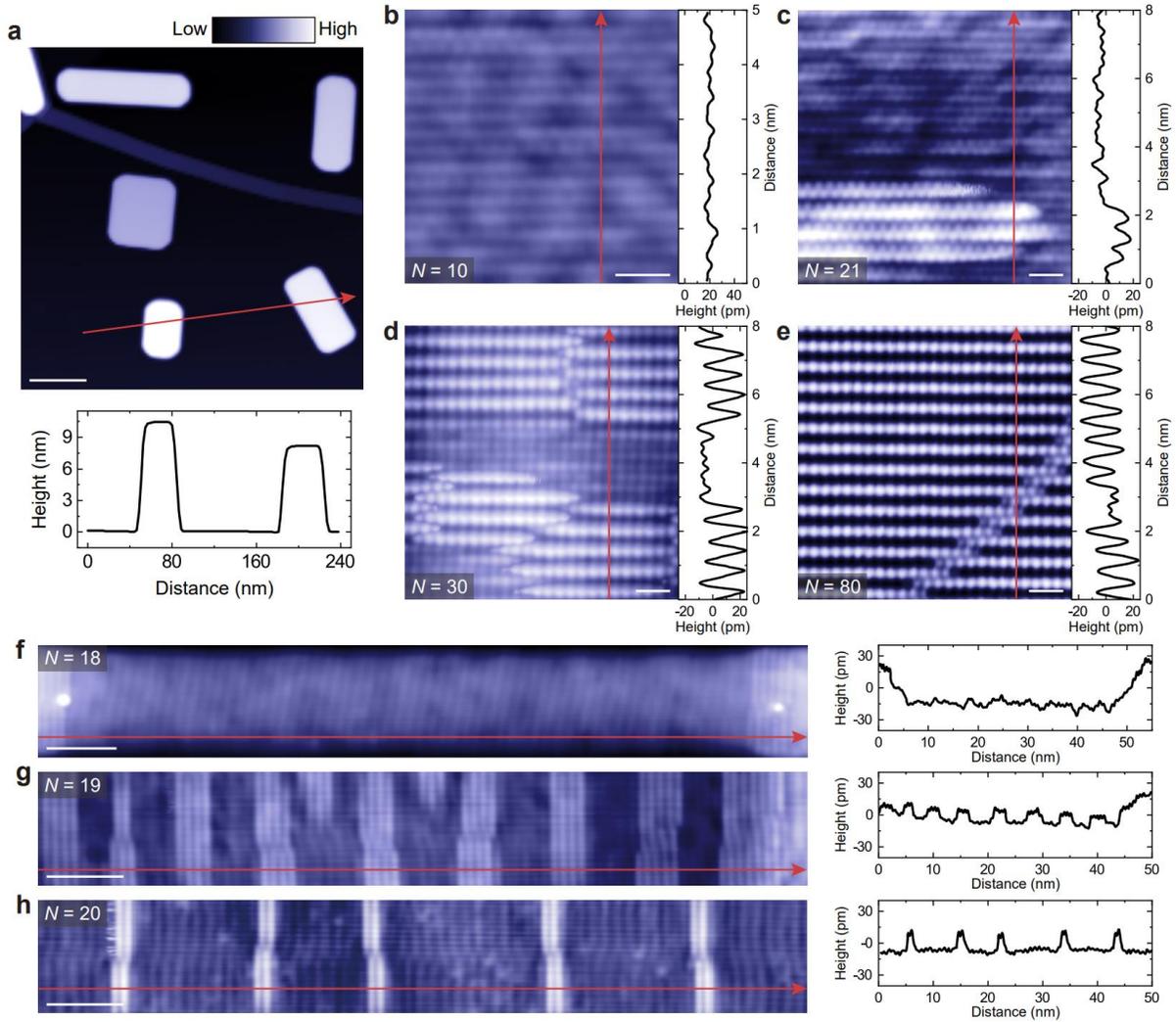

**Fig. 2| Surface morphology of β-Sn(100) islands with different *N*. a,** Large scale STM image (300 × 300 nm$^2$) of β-Sn(100) islands on graphene-terminated 6H-SiC(0001) (sample bias $V_B$ = 1.5 V and tunneling current $I_t$ = 0.02 nA). **b-e,** Atomic resolution STM images of β-Sn(100) islands with *N* = 10 (**b**), *N* = 21 (**c**), *N* = 30 (**d**), and *N* = 80 (**e**). STM setpoints in (**b-e**): $V_B$ = 10 mV and $I_t$ = 8 nA. **f-h,** STM images of β-Sn(100) islands with *N* = 18 (**f**), *N* = 19 (**g**), and *N* = 20 (**h**). STM setpoints in (**f-h**): $V_B$ = 100 mV and $I_t$ = 1 nA. The bottom panel in (**a**) and the right panels in (**b-h**) show height profiles along the red arrows in the corresponding STM images. Scale bars: 50 nm (**a**); 1 nm (**b-e**); 5 nm (**f-h**). All STM measurements are performed at *T* ~ 4.2 K.

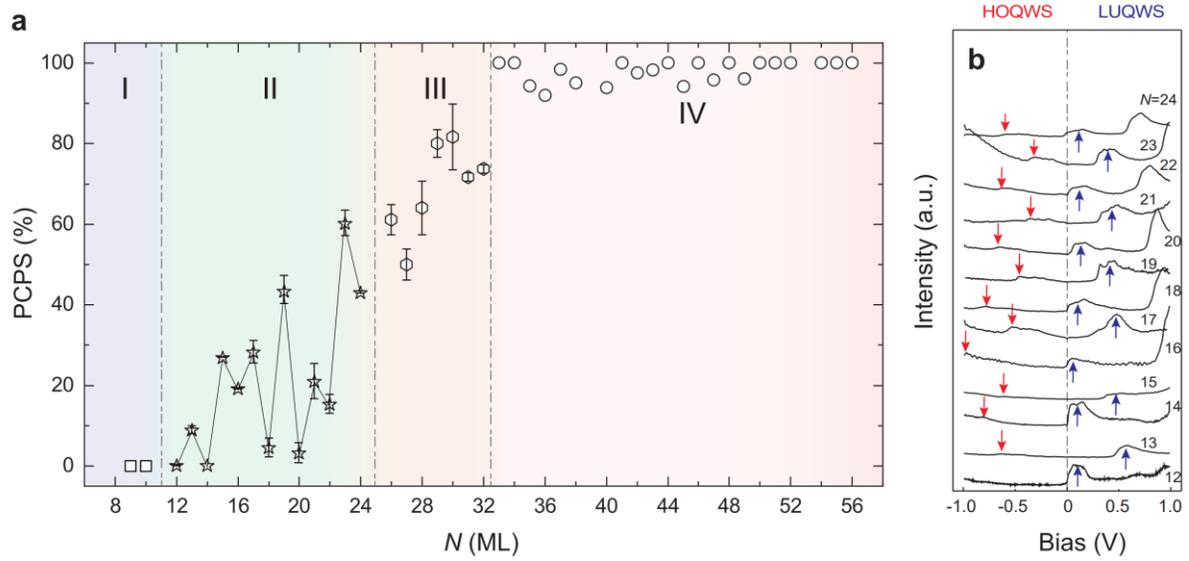

**Fig. 3| $N$ dependence of PCPS and QW states in β-Sn(100) islands. a,** PCPS as a function of $N$. For $12 \leq N \leq 32$, the error bar is calculated as the standard deviation of PCPS from different samples with the same $N$. **b,** d$I$/d$V$ spectra on β-Sn(100) islands with $12 \leq N \leq 24$, i.e., Range II in (**a**). The highest occupied QW states (HOQWS) and the lowest unoccupied QW states (LUQWS) are marked by red and blue arrows, respectively. The dashed black line represents the Fermi level $E_F$.

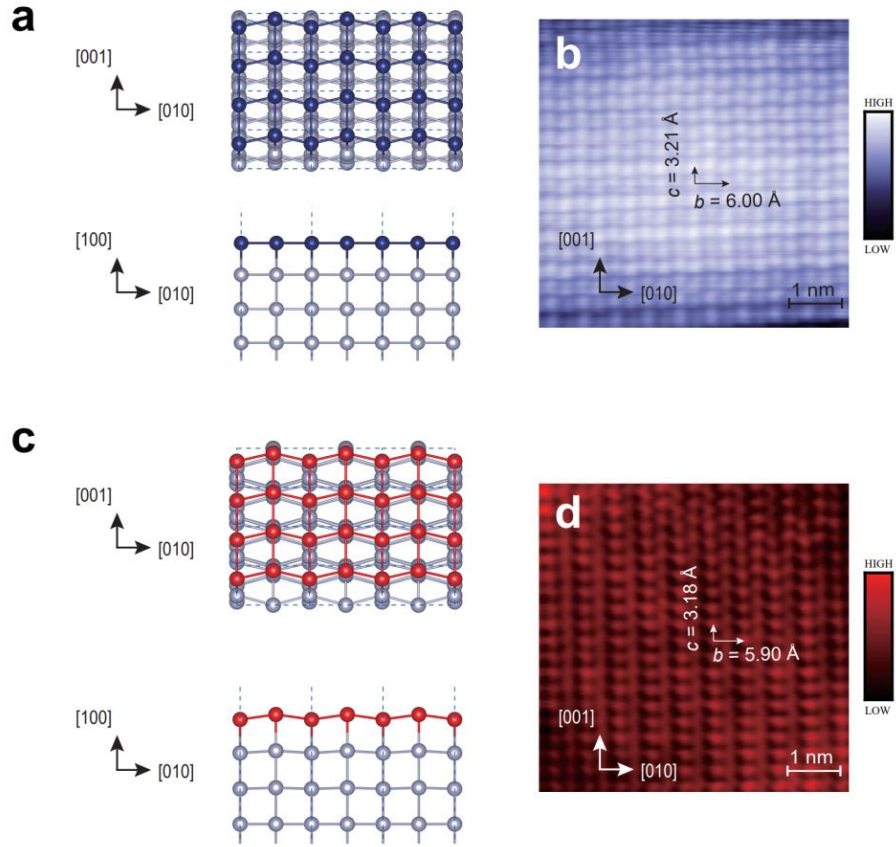

**Fig. 4| Flat and patterned surfaces on β-Sn(100) islands. a, c,** Lattice structures of flat (**a**) and patterned (**c**) surfaces. **b, d,** Atomic-resolution STM images of flat (**b**, $V_B$ =100 mV and $I_t$ =0.3 nA) and patterned (**d**, $V_B$ =20 mV and $I_t$ =0.3 nA) surfaces. All STM measurements are performed at $T$ ~ 4.2 K.

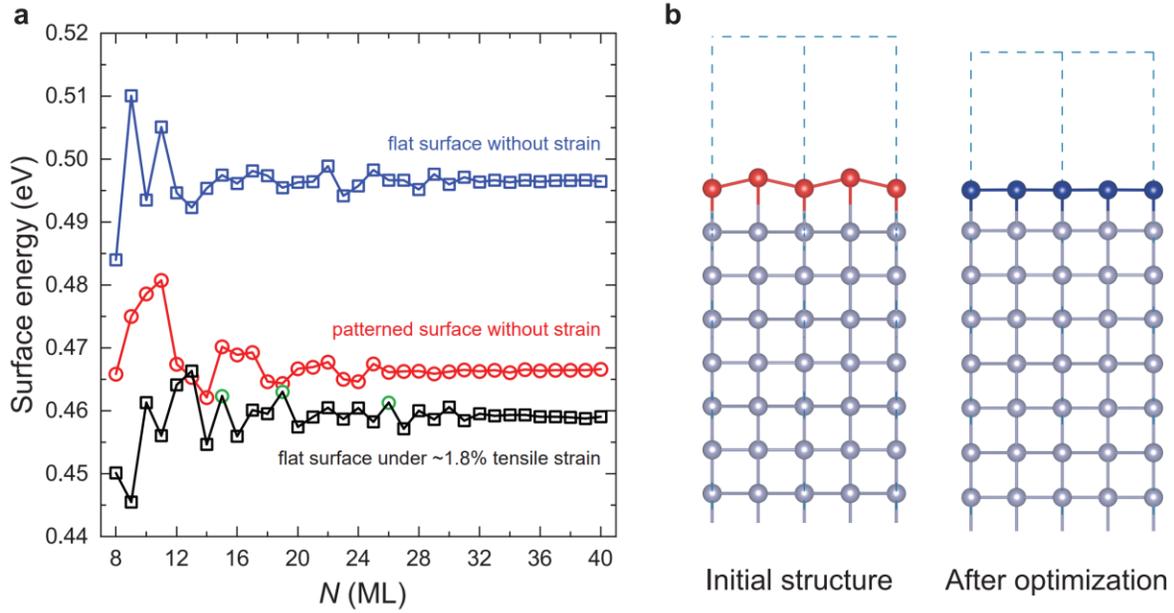

**Fig. 5| Surface energies and lattice structure modeling of β-Sn(100) films. a**, Calculated surface energies of flat and patterned β-Sn(100) films as a function of $N$ at equilibrium, together with the surface energies of flat β-Sn(100) films under ~1.8% biaxial tensile strain. For the strained case, films with $N$ = 15, 19, and 26 (highlighted by green circles) are initially prepared in flat configurations but relaxed into patterned surfaces, indicating that QSE influences the relative stability of flat and patterned surfaces at intermediate thicknesses under this strain level, consistent with our experimental observations in Range II. **b**, Schematics of an initially patterned β-Sn(100) surface relaxing into a flat surface under sufficiently large biaxial tensile strain.

.